\documentclass[10pt,conference]{IEEEtran}
\IEEEoverridecommandlockouts
\usepackage{cite}
\usepackage{amsmath,amssymb,amsfonts}
\usepackage{algorithmic}
\usepackage{graphicx}
\usepackage{textcomp}
\usepackage{xcolor}
\usepackage{indentfirst}
\usepackage{hyphenat}
\usepackage{booktabs}
\usepackage{listings}
\usepackage{url}

\def\BibTeX{{\rm B\kern-.05em{\sc i\kern-.025em b}\kern-.08em
    T\kern-.1667em\lower.7ex\hbox{E}\kern-.125emX}}

\begin{document}

\title{Joule-Profiler: Profiling the Energy Consumption of Build Automation Tools Made Easy}

\author{
\IEEEauthorblockN{Jérémy Woirhaye\IEEEauthorrefmark{1}}
\IEEEauthorblockA{Inria / Univ.\,Lille / CNRS, CRIStAL\\France\\jeremy.woirhaye@inria.fr}
\and
\IEEEauthorblockN{François Gibier\IEEEauthorrefmark{1}}
\IEEEauthorblockA{Inria / Univ.\,Lille / CNRS, CRIStAL\\France\\francois.gibier@inria.fr}
\and
\IEEEauthorblockN{Romain Rouvoy}
\IEEEauthorblockA{Inria / Univ.\,Lille / CNRS, CRIStAL\\France\\romain.rouvoy@inria.fr}
\thanks{\IEEEauthorrefmark{1}Both authors contributed equally to this work.}
}

\maketitle

\begin{abstract}
Build pipelines are integral to modern software development, yet their energy footprint remains largely invisible to practitioners.
Existing CI energy tools either rely on model-based estimation (due to hardware access restrictions in cloud runners) or report only total pipeline energy without decomposing it into meaningful phases.
{\sf Joule-Profiler} is an open-source command-line tool for Linux that measures hardware energy consumption via Intel RAPL (CPU), NVML (NVIDIA GPU), and attributes it to user-defined program phases by monitoring standard output.
In this tool paper, we demonstrate {\sf Joule-Profiler} in the context of Maven build pipelines using Google Gson as a case study.
By applying a token pattern-matching Maven plugin invocations, we analyze builds across the 5 most recent Gson releases into per-phase energy profiles, comparing cold builds (empty local repository) and warm builds (cached dependencies).

A screencast demonstrating Joule Profiler is available at: \url{https://www.youtube.com/watch?v=v64Lactu5HQ}, the source code is available at: \url{https://github.com/joule-profiler/joule-profiler}, and the reproduction package, including the Nix flake, scripts, raw data, and analysis notebooks, is available at: \url{https://github.com/joule-profiler/reproduction-package-icse-2027}
\end{abstract}

\begin{IEEEkeywords}
Energy profiling, Phase-based measurement, RAPL, Maven, CI/CD, Build pipelines, Joule Profiler
\end{IEEEkeywords}

\section{Introduction}
Software energy consumption has emerged as a first-class concern for both practitioners and researchers. CI/CD pipelines run continuously, triggered by every commit, pull request, or scheduled job, and their cumulative energy cost across large projects and organizations is non-negligible~\cite{perez2024software,de2025evaluating}.
Understanding where energy is spent within a pipeline execution is a prerequisite for targeted optimization.

Existing approaches fall into two categories.
Model-based tools, such as {\sf Eco\,CI}~\cite{ecocigithub}, estimate energy from CPU utilization metrics and are compatible with cloud-hosted runners, but are limited to estimations and cannot attribute energy to fine-grained phases within a single pipeline step.
Direct-measurement tools, such as {\sf PowerAPI}~\cite{powerapi}, {\sf Alumet}~\cite{alumet}, {\sf Scaphandre}~\cite{scaphandre}, and {\sf EnergiBridge}~\cite{energibridge}, provide hardware-accurate readings but are designed for system- or container-level monitoring rather than intra-process phase decomposition.

{\sf Joule-Profiler}~\cite{Woirhaye_Joule_Profiler_A_2026} fills this gap by combining hardware-accurate measurement (Intel RAPL via \texttt{perf\_event}) with a lightweight phase-detection mechanism. It monitors the target process's standard output and matches lines against a user-defined regular expression to mark phase boundaries. No instrumentation library or code modification is required.

In this paper, we showcase {\sf Joule-Profiler} on a concrete software engineering use case: profiling the energy consumption of Maven builds across the release history of Google Gson~\cite{gson}.
This scenario is representative of a broad class of build-system workflows and illustrates how {\sf Joule-Profiler} can support energy regression in projects, detecting phases that become more energy-intensive as a project evolves, a workflow not supported by existing tools.

\section{Background and Related Work}

{\sf Eco\,CI}~\cite{ecocigithub} is a GitHub Action and GitLab plugin easily integrated into existing workflows that estimates energy consumption of CI pipeline steps by sampling CPU utilization and applying a machine-learning power model derived from the SPECpower database.
Its fundamental limitation is that it provides only estimates, not measurements, because RAPL or other hardware counters are unavailable in virtualized cloud runners.
Moreover, it reports energy per pipeline step, a unit defined by the CI configuration, rather than per program phase, a semantic unit defined by the program's execution structure.
Thus, within a single \texttt{mvn install} step, {\sf Eco\,CI} cannot distinguish the energy spent downloading dependencies, compiling sources, running tests, or packaging the artifact.
{\sf PowerAPI}~\cite{powerapi}, {\sf Alumet}~\cite{alumet}, and {\sf Scaphandre}~\cite{scaphandre} are daemon-based infrastructure monitoring tools designed for continuous observability across fleets of machines or containers.
They are not designed for single-invocation, experimental workflow use cases.
{\sf EnergiBridge}~\cite{energibridge} delivers cross-platform energy measurement as a command-line wrapper, but without phase decomposition.
Finally, {\sf JouleIt}~\cite{jouleit}, the direct predecessor of {\sf Joule-Profiler}, demonstrated a lightweight wrapper approach but lacked support for phases and GPU measurements.

\section{Tool Overview}

{\sf Joule-Profiler} is invoked as a command-line wrapper:
\begin{lstlisting}[language=bash]
joule-profiler profile --token-pattern "<regex>" -- <command>
\end{lstlisting}

It spawns the target command as a child process, continuously reads its standard output, and matches each line against the provided regular expression. Each match delimits a new phase.
At each phase boundary, {\sf Joule-Profiler} snapshots the current RAPL energy counters ({\tt PACKAGE}, {\tt DRAM}, and other available domains). After the process exit, it computes phase energy as the difference between consecutive snapshots and reports results to the terminal or exports them in JSON or CSV format.

The tool accesses RAPL via the \texttt{perf\_event} interface and falls back to the Linux \texttt{powercap} sysfs interface if \texttt{perf\_event} is unavailable. Sources run as asynchronous tasks to minimize measurement overhead.
Prior validation demonstrated that {\sf Joule-Profiler} measurements are statistically equivalent to {\sf perf} for RAPL domains and to {\sf Alumet}~\cite{alumet} for GPU (Pearson correlation of 99.9\% for both {\tt PACKAGE} and {\tt DRAM}, and 99.5\% for GPU), with a mean phase-detection delay below 36~$\mu$s, well within the 1~ms RAPL counter refresh rate~\cite{Woirhaye_Joule_Profiler_A_2026}.

\section{Case Study: Energy Profiling of Maven Builds}

\subsection{Motivation}

Maven is one of the most widely used build systems in the Java ecosystem~\cite{perez2024software}. A \texttt{mvn install} invocation encompasses multiple sequential phases, each driven by a distinct Maven plugin, including dependency resolution, resource filtering, compilation, test execution, and packaging. These phases have very different energy profiles. Dependency resolution is I/O-bound and network-bound on \emph{cold} builds, while compilation and testing are CPU-bound. Tracking per-phase energy across a project's release history enables practitioners to identify energy regressions, for example, a new release that significantly increases test execution energy due to added tests or slower algorithms. Phase-level attribution makes this identification unambiguous, whereas total-energy tools would obscure which phase regressed.

\subsection{Experimental Protocol}

We consider Google Gson~\cite{gson} as our subject project and select its 5 most recent releases. For each release, we apply the following protocol, repeated 40 times per configuration, to reduce measurement variability.

\noindent\textbf{Step 1 -- \emph{Cold} build.}
We empty the local Maven repository (\texttt{\textasciitilde/.m2/repository}) and execute:

\begin{lstlisting}[language=bash]
joule-profiler profile --token-pattern "(?P<plugin>[\w-]+):(?P<version>[\d.]+):(?P<goal>[\w-]+)\s+\((?P<id>[^)]+)\)\s+@\s+(?P<module>[\w-]+)" -- mvn install
\end{lstlisting}

\noindent This configuration instructs {\sf Joule-Profiler} to create a new phase each time Maven logs a plugin invocation to stdout (e.g., \texttt{maven-compiler-plugin:3.11.0:compile (default-compile) @ module-name}), capturing both the plugin execution and the target module. Phases thus correspond directly to Maven lifecycle plugin executions across all submodules, covering each build step.

\noindent\textbf{Step 2 -- Build cleanup.}
Execute \texttt{mvn clean} to remove build artefacts while retaining the populated local repository.

\noindent\textbf{Step 3 -- \emph{Warm} build.}
Re-execute the identical Joule Profiler command. Since all dependencies are already cached in \texttt{\textasciitilde/.m2/repository}, this measurement isolates the computational cost of compilation and testing from network and I/O download costs.

\noindent\textbf{Step 4 -- Reset.}
Execute \texttt{mvn clean} and purge downloaded dependencies from \texttt{\textasciitilde/.m2}.

This protocol yields, for each release, a cold build profile showing per plugin energy, including dependency resolution, and a warm build profile isolating pure build energy. Comparing the two directly quantifies the energy cost of dependency resolution. Comparing warm builds across releases reveals energy regressions in compilation or testing.

For these experiments, we used Java 17 and Maven 3.9.11. To make the results as reproducible as possible, we fixed these versions using Nix, a reproducible package manager provided in the reproduction package. All runs were performed on a single \textit{chirop} node of Grid'5000 (2 Intel Xeon Platinum 8358, 512~GB RAM, kernel version 5.10.0-38-amd64), with 1~s cool-down between runs.

\subsection{Experiments Results}

\begin{figure}[htbp]
  \centering
  \includegraphics[width=\linewidth]{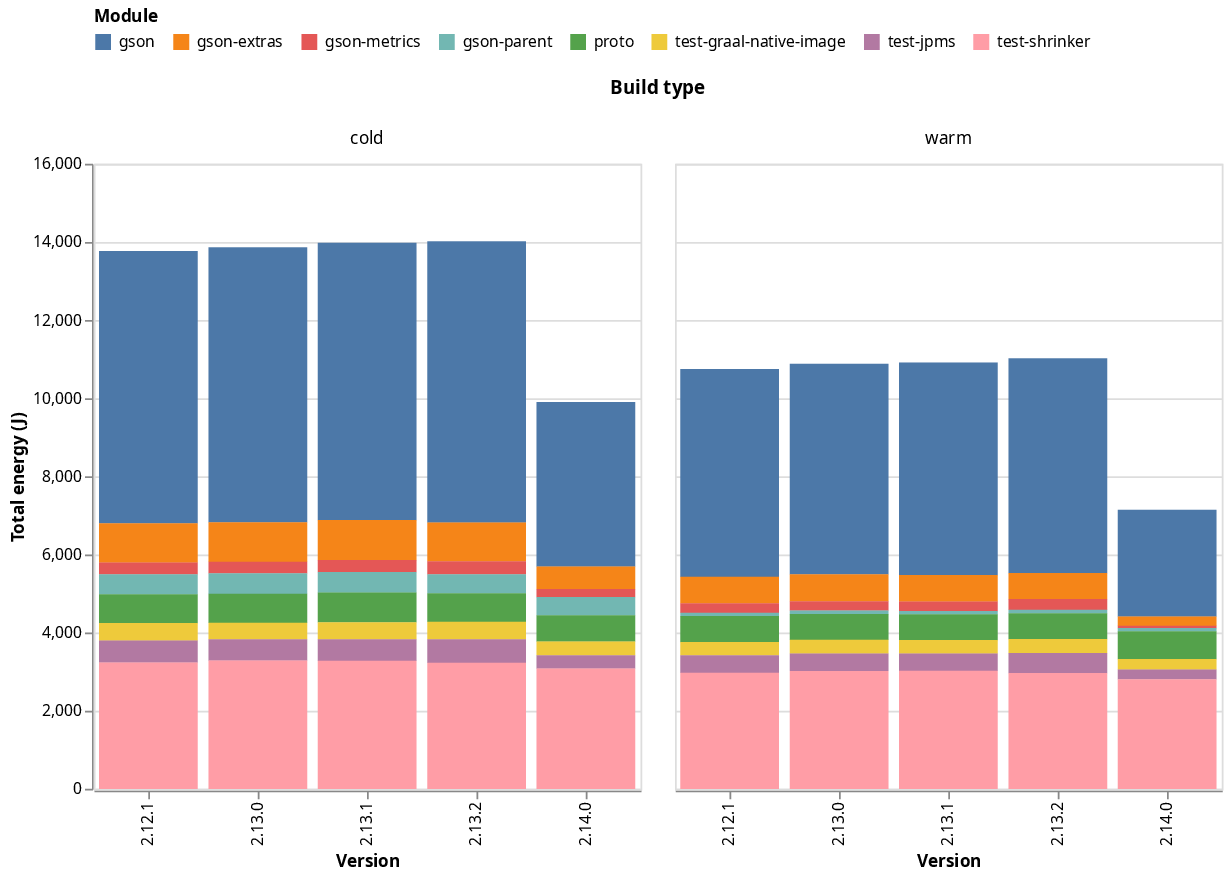}
  \caption{Total energy (J) per release and per module for cold and warm builds. Each bar represents a release, stacked by the Maven Gson submodule.}
  \label{fig:bars}
\end{figure}

\textbf{Cold vs warm comparison}: Fig.~\ref{fig:bars} shows the total energy consumption per release, broken down by module, for both cold and warm builds. Each bar represents a release, and each stack corresponds to a submodule of the Maven project Gson~\cite{gson} (gson, gson-extras, gson-metrics, gson-parent, proto, test-graal-native-image, test-jpms, test-shrinker). Energy consumption differs visibly between cold and warm builds. The cold/warm ratio ranges from 1.28 (release~2.12.1) to 1.386 (release~2.14.0), reflecting the impact of dependency resolution in cold builds. This highlights the benefit of implementing caching mechanisms, especially since a pipeline can run many times per day, thereby reducing energy consumption~\cite{perez2024software,11500151}.

\textbf{Cross-release energy evolution}:
Fig.~\ref{fig:bars} shows that energy consumption stays stable across versions~2.12.1 to~2.13.2 for both cold and warm builds (between 13,760~J and 14,033~J for cold, and between 10,746~J and 11,021~J for warm), before dropping sharply in~2.14.0. The \textit{gson} submodule is the main driver of this drop, accounting for around 51.4\% of total cold energy in version~2.13.2 (7,208~J out of 14,033~J). It went from 7,208~J to 4,211~J in cold builds (41.6\% less) and from 5,497~J to 2,729~J in warm builds (50.4\% less). Other submodules also decreased between versions~2.13.2 and~2.14.0, notably \textit{gson-metrics} (77.6\% in warm) and \textit{test-jpms} (51.5\% in warm), while the remaining modules stayed relatively stable, confirming that the changes in version~2.14.0 mainly affected the \textit{gson} submodule and to a lesser extent \textit{gson-metrics} and \textit{test-jpms}.

\begin{figure}[htbp]
  \centering
  \includegraphics[width=\linewidth]{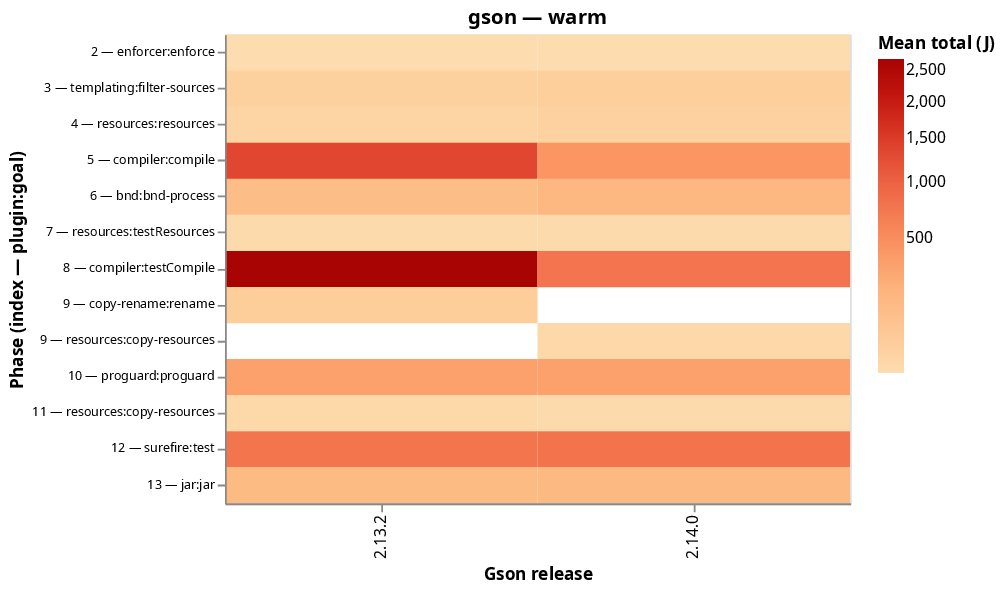}
  \caption{Per-phase mean energy (J) heatmap for the \textit{gson} submodule across the 2.13.2 and 2.14.0 releases for warm builds. Each row is a Maven plugin invocation identified by its index and \texttt{plugin:goal} name.}
  \label{fig:heatmap_gson}
\end{figure}

\textbf{Maven goal energy evolution in the gson sub-project}:
Fig.~\ref{fig:heatmap_gson} shows the mean energy of each phase of the \textit{gson} submodule for warm builds across the 2.13.2 and 2.14.0 releases. Two phases explain almost all the drop in 2.14.0: \texttt{compiler:compile} (phase~5) and \texttt{compiler:testCompile} (phase~8). Both stay roughly constant from 2.12.1 to 2.13.2 and then drop sharply in 2.14.0, while every other phase, including \texttt{bnd:bnd-process}, \texttt{proguard:proguard}, and \texttt{sure\hyp{}fire:test}, stays stable. The swap between \texttt{copy-rename:rename} and \texttt{resour\hyp{}ces:copy-resources} at index~9 reflects a build configuration change, not an energy regression. The drop is therefore concentrated in the two Java compilation steps.

A first hypothesis would be that the codebase shrank between the two releases. Table~\ref{tab:gson_codebase} reports lines of code and test counts for each submodule. The differences are small and cannot explain a drop of more than 70\% on the compilation phases. The energy of \texttt{surefire:test} is also stable across all releases, meaning the tests run the same amount of work as before. 
The drop therefore does not come from changes in the code being compiled or executed, but from a change in how compilation is performed.

\begin{table}[htbp]
  \centering
  \caption{Lines of code and test methods per submodule in versions~2.13.2 and~2.14.0 (measured with SonarQube).}
  \label{tab:gson_codebase}
  \begin{tabular}{lrrrr}
    \toprule
    Submodules & \multicolumn{2}{c}{Lines of code} & \multicolumn{2}{c}{Test methods} \\
    \cmidrule(lr){2-3}\cmidrule(lr){4-5}
    \multicolumn{1}{r}{Versions} & 2.13.2 & 2.14.0 & 2.13.2 & 2.14.0 \\
    \midrule
    \texttt{gson}                    & 10{,}160 & 10{,}689 & 4{,}515 & 4{,}545 \\
    \texttt{gson-extras}             & 761 & 542 & 40 & 30 \\
    \texttt{gson-metrics}            & 773 & 759 & 0 & 0 \\
    \texttt{proto}                   & 362 & 897 & 21 & 148 \\
    \texttt{test-graal-native-image} & 133 & 124 & 10 & 10 \\
    \texttt{test-jpms}               & 47  & 33  & 12 & 12 \\
    \texttt{test-shrinker}           & 789 & 779 & 0 & 0 \\
    \texttt{pom.xml} (parent file)   & 505 & 509 & 0 & 0 \\
    \bottomrule
  \end{tabular}
\end{table}

This pattern points at Error Prone, a static analysis tool that runs as a \texttt{javac} annotation processor and therefore inflates \texttt{compiler:compile} and \texttt{compiler:testCompile} without affecting any other goal. Both releases declare \texttt{error\_prone\_core} as a compiler dependency (2.41.0 in 2.13.2, 2.48.0 in 2.14.0) and define a \texttt{disable-error-prone} Maven profile that disables it within a JDK range. Only the activation range differs:

\begin{lstlisting}[language=XML,caption={Activation in 2.13.2}]
<profile>
  <id>disable-error-prone</id>
  <activation><jdk>[,17)</jdk></activation>
  ...
</profile>
\end{lstlisting}

\begin{lstlisting}[language=XML,caption={Activation in 2.14.0}]
<profile>
  <id>disable-error-prone</id>
  <activation><jdk>[,21)</jdk></activation>
  ...
</profile>
\end{lstlisting}

Our experiments use Java~17. In 2.13.2, the range \texttt{[,17)} excludes Java~17, so the profile does not run and Error Prone is active. In 2.14.0, the range is widened to \texttt{[,21)}, which now includes Java~17: the profile runs and fully disables Error Prone even though the dependency is still declared.

To confirm this single change explains the drop, we ran a control experiment on 2.14.0: we downgraded Error Prone to 2.41.0 and restored the activation range to \texttt{[,17)}. The energy of \texttt{compiler:compile} and \texttt{compiler:testCompile} returns to the 2.13.2 level while all other phases stay unchanged. The same downward trend appears across the other Java submodules, since the change was made in the parent \texttt{pom.xml} and therefore propagates to every module in the project. This shows that configuration files such as \texttt{pom.xml}, which are rarely modified compared to source code, can significantly affect a build's energy footprint, and that energy profiling should not be limited to source code changes.

This phase-level attribution directly demonstrates the energy regression detection capability of {\sf Joule-Profiler}: a developer observing a 50.3\% drop in warm build energy for the \textit{gson} submodule between~2.13.2 and~2.14.0 can immediately identify, from Fig.~\ref{fig:heatmap_gson}, that the reduction is concentrated in \texttt{compiler:compile} and \texttt{compiler:testCompile} rather than in other build phases, enabling a targeted and actionable investigation that recovers the root cause from a few lines of \texttt{pom.xml}.

\section{Discussion}
\subsection{Beyond Maven and Gson}
Although our case study focuses on Maven, {\sf Joule-Profiler} 's stdout-driven phase-detection mechanism is build-system agnostic. Any tool that emits a deterministic, line-oriented lifecycle log can be profiled by adjusting the token pattern. The same approach applies, for instance, to Gradle, npm lifecycle hooks and Cargo. Beyond build systems, the same mechanism can attribute energy to a test suite or to stages in a data pipeline, enabling fine-grained energy regression testing in domains well beyond CI build energy.

\subsection{Positioning against existing tools}
{\sf Eco\,CI}~\cite{ecocigithub} is the most widely used tool for tracking CI energy today.
It works on cloud-hosted runners because it does not rely on hardware counters: since RAPL is not available in virtual machines, {\sf Eco\,CI} estimates energy from CPU usage and a power model.
This design has two key limitations.
On bare-metal or self-hosted runners, which are common in research and in regulated industries, direct RAPL measurement is possible and gives more accurate results.
More importantly, {\sf Eco\,CI} reports energy per CI step, so a single \texttt{mvn install} is treated as a black box, whereas {\sf Joule-Profiler} lets developers choose the level of detail they want by writing a regular expression. Other tools such as PowerAPI~\cite{powerapi}, Alumet~\cite{alumet}, and Scaphandre~\cite{scaphandre} are not direct competitors: they monitor whole machines or fleets of servers over long periods, not a single command run by run. EnergiBridge~\cite{energibridge} and JouleIt~\cite{jouleit} work as command-line wrappers like {\sf Joule-Profiler}, but neither of them supports phase decomposition.
The two approaches can also be used together: {\sf Eco\,CI} gives a broad picture of energy trends across many cloud pipeline runs, while {\sf Joule-Profiler} gives precise per-plugin measurements on dedicated hardware, which can then guide targeted optimizations.

\subsection{Limitations and threats to validity}
{\sf Joule-Profiler} relies on three assumptions.
First, phases are detected from standard output, so the target program must log each step on its own line in a predictable order.
Programs that print logs in bursts will produce less accurate phase boundaries because hardware counters, such as RAPL, have a limited refresh rate.
Second, {\sf Joule-Profiler} currently measures the energy used by the entire machine for RAPL, not just the target program. RAPL reports the energy of the entire CPU package, including all concurrently running processes.
The values we report, therefore, constitute an upper bound on the program's actual consumption.
To get closer to the program's actual cost, one must subtract the energy used by the machine when idle and keep other activities to a minimum, as we did by running each build on a dedicated Grid'5000 node.

\section{Conclusion}
We demonstrate {\sf Joule-Profiler} on Maven build pipelines, using Google Gson~\cite{gson} as a case study across 5 releases.
{\sf Joule-Profiler}'s stdout-based phase detection, combined with direct RAPL measurement, provides per-plugin energy profiles that existing CI energy tools cannot capture.
The cold/warm build comparison protocol isolates dependency resolution costs from computational costs, while cross-release tracking of warm build profiles enables energy regression detection at plugin granularity.
This capability positions {\sf Joule-Profiler} as a practical tool for energy-aware software engineering in environments with direct hardware access, complementing model-based approaches like {\sf Eco\,CI} in cloud settings.

\section*{Acknowledgment}
This work received support from the France 2030 program, managed by the French National Research Agency under grant agreement No. ANR-23-PECL-0003 (PEPR Cloud CARECloud), and from the Inria/Qarnot PULSE project (\url{https://defi-pulse.github.io/}).
Experiments were carried out using the Grid'5000 testbed, supported by a scientific interest group hosted by Inria and including CNRS, RENATER, and several Universities (\url{https://www.grid5000.fr}).

\bibliographystyle{IEEEtran}
\bibliography{sample-base}

\end{document}